\documentclass[prl,aps,nofootinbib,superscriptaddress,tightenlines,twocolumn]{revtex4}

\usepackage{epsfig}
\usepackage{color}
\usepackage[latin1]{inputenc}
\usepackage{float,amsmath,slashed}
\usepackage{graphicx}

\begin{document}

\title{The evolving distribution of hard partons traversing a hot strongly interacting plasma}

\author{Bj\"orn Schenke}
\affiliation{Department of Physics, McGill University, Montreal, Quebec, H3A\,2T8, Canada}

\author{Charles Gale}
\affiliation{Department of Physics, McGill University, Montreal, Quebec, H3A\,2T8, Canada}

\author{Guang-You Qin}
\affiliation{Department of Physics, The Ohio State University, Columbus, OH 43210, USA}

\begin{abstract}
We present results for the evolution of the momentum distribution of a hard parton traversing a brick of quark-gluon plasma, considering both bremsstrahlung and collisional energy loss. The complete leading order transition rates are included. We find a significant modification of the evolving momentum distribution compared to results obtained by employing an approximate implementation of the collisional energy loss that uses the mean energy loss, together with momentum diffusion.
\end{abstract}

\maketitle

\section{Introduction}
High transverse momentum jets emerging from the central rapidity region
in heavy ion collisions can provide important information on the created hot quark-gluon plasma (QGP).
After the discovery of jet quenching at the Relativistic Heavy Ion Collider 
\cite{Adcox:2001jp,Adler:2002xw} a lot of progress has been made toward using jets as a quantitative tomographic probe of the QGP
\cite{Gyulassy:1993hr,Baier:1996sk,Baier:1998yf,Zakharov:2000iz,Wang:2001if,Vitev:2002pf,Jeon:2003gi,Salgado:2003gb,Majumder:2004pt,Wicks:2005gt,Zhang:2007ja,Qin:2007zz,Qin:2007rn,Fochler:2008ts,Schenke:2008gg}.
Jet quenching refers to the suppression of high transverse momentum hadrons, such as $\pi^0$ and $\eta$ mesons in central $Au+Au$ collisions compared to expectations from measurements in $p+p$ collisions.
This suppression has been attributed to the energy loss of hard $p_T$ partons due to induced gluon
bremsstrahlung in the hot quark-gluon plasma phase.  
Several theoretical formalisms have been established to describe the energy loss due to bremsstrahlung
\cite{Baier:1996kr,Gyulassy:2000er,Kovner:2003zj,Zakharov:2007pj,Wang:2001if,Arnold:2001ms}.

In addition, collisional energy loss should play a role in the suppression of high momentum jets. However, its importance compared to radiative losses has been under discussion.
On the one hand, there have been arguments for that its contribution is small compared to radiative energy loss.
For example, early estimates using asymptotic arguments indicated that the radiative energy loss is much larger than
the elastic energy loss \cite{Bjorken:1982tu}.
Furthermore, in \cite{Zakharov:2007pj} radiative energy loss in the light-cone path integral
approach and collisional energy loss employing the Bjorken method were compared and
collisional energy loss was found to be smaller.
In \cite{Renk:2007id} phenomenological limits on radiative vs. collisional energy loss were derived by considering quadratic vs. linear pathlength dependence, again finding that any elastic energy loss component has to be small.
On the other hand, in \cite{Mustafa:2003vh,Mustafa:2004dr} it was found that
collisional energy loss has a significant influence on jet quenching, and recent studies \cite{Adil:2006ei,Wicks:2007mk} also point in this direction. See in addition Refs. \cite{Majumder:2008zg} and \cite{Wang:2006qr}.

In several works \cite{Wicks:2005gt,Qin:2007rn} the evolution of the jet momentum distribution including both radiative and collisional energy loss has been computed. In both works it was found that, although the total energy loss of light quarks is dominated by radiative processes, the shape of the momentum distribution depends significantly on whether elastic processes are included or not. However, motivated by the fact that collisional energy loss is dominated by small momentum transfers, in both \cite{Wicks:2005gt} and \cite{Qin:2007rn} elastic collisions were approximated by a mean energy loss, accompanied by momentum diffusion. 

Because it was found that the evolving shape of the jet momentum distribution can be strongly affected by including elastic processes, it is important to go beyond this approximation, which from here on we refer to as diffusion method. In \cite{Wicks:2007mk} modifications to this method due to fluctuations have been investigated and found to be important. 
In this work we include the full perturbative transition rates to obtain a more precise result and test the validity of the approximation. We simulate the situation of a high momentum parton traversing a brick of static QGP, considering collisional and radiative energy loss \cite{Arnold:2001ba,Arnold:2001ms,Arnold:2002ja,Jeon:2003gi} separately and combined. In both cases we study the evolution of the parton's entire momentum distribution, comparing the diffusion method to the direct leading order calculation.

\section{Formalism}
The jet momentum distribution $P(E,t)=dN(E,t)/dE$ evolves
in the medium according to a set of coupled Fokker-Planck type rate equations of the 
form \cite{Qin:2007zz}:
\begin{align}\label{jet-evolution-eq}
\frac{dP(E)}{dt}\!=\!\int_{-\infty}^{\infty}\!\!\!\!\!\!d\omega
\left(\!P(E{+}\omega) \frac{d\Gamma(E{+}\omega,\omega)}{d\omega} - P(E)\frac{d\Gamma(E,\omega)}{d\omega}\!\right)
\end{align}
where ${d\Gamma(E,\omega)}/{d\omega}$ is the transition rate for
processes where partons of energy $E$ lose energy $\omega$. The
$\omega<0$ part of the integration incorporates processes which increase
a particle's energy.  The radiative part of the transition rate is
discussed in \cite{Jeon:2003gi, Turbide:2005fk, Qin:2007zz} and the elastic part is discussed in detail in the following.

First, we briefly review the transition rates of the diffusion method used in \cite{Qin:2007rn}. It was argued in this work that it should be an adequate procedure to approximate transition rates due to elastic collisions of the hard parton with the thermal medium using the mean energy loss (drag term) and a momentum diffusion term as dictated by detailed balance.
The motivation for this was that compared to radiative energy loss, collisional losses are more dominated by small energy transfers because the contribution to the mean energy loss rate $dE/dt$ from elastic collisions is only 
logarithmically sensitive to large energy transfers. On the other hand, the radiative contribution is a power-law, for large values of the radiated energy. It was thus anticipated that, as long as radiative energy losses dominate jet quenching, this method will adequately address the effects of elastic collisions. We investigate this assertion quantitatively in this work by going beyond the diffusion approximation and  monitoring  the entire profile of the energy spectrum.

In practice, the analytic results for the energy loss rate $dE/dt$ \cite{Braaten:1991jj,Braaten:1991we,Thomas:1991ea,GaleKapusta:1992}
were incorporated into (\ref{jet-evolution-eq}) by introducing the drag term, $(dE/dt)dP(E)/dE$, and
the diffusion term, $T(dE/dt)d^2P(E)/dE^2$.  Then Eq.~(\ref{jet-evolution-eq})
was discretized, such that $\int d\omega \rightarrow \Delta\omega
\sum_{\omega = n\Delta \omega}$, and 
\begin{eqnarray}
\Gamma(E+\Delta\omega,\Delta\omega) &=& (1+f_B(\Delta\omega))(\Delta
\omega)^{-1} dE/dt \, , \nonumber \\
\Gamma(E,-\Delta \omega) & = & f_B(\Delta\omega) (\Delta\omega)^{-1}
dE/dt \,,
\end{eqnarray}
which yields the right energy loss rate and preserves detailed balance, for small enough $\Delta\omega$. 
Using this method the momentum distribution evolves like a Gaussian with increasing width, shifted by the mean energy loss at the regarded time.

Next, two methods for computing the transition rates beyond the diffusion approximation are discussed. One may start from the expression for the transition rate  
\begin{align}\label{tr}
 \frac{d\Gamma}{d\omega}~(E,\omega,T)=&d_k\int \frac{d^3k}{(2\pi)^3}\int \frac{d^3k^\prime}{(2\pi)^3}\frac{2\pi}{16pp^\prime k k^\prime}\notag\\ &\times\delta(p-p^\prime-\omega)\delta(k^\prime-k-\omega)\notag\\
 &\times|\mathcal{M}|^2 f(k,T)(1\pm f(k^\prime,T))\,,
\end{align}
where $p=E=|\mathbf{p}|$ and $p^\prime=|\mathbf{p}^\prime|$ are the absolute values of the three-momenta of the incoming and outgoing hard parton, respectively, and $k=|\mathbf{k}|$ and $k^\prime=|\mathbf{k}^\prime|$ are those of the incoming and outgoing thermal parton. In addition, $\omega=p-p^\prime=k^\prime-k$ is the transferred energy.
The distribution functions $f$ are either Fermi-Dirac or Bose-Einstein distributions depending on the nature of the thermal parton involved. The $+$ or $-$ sign appears accordingly, with $-$ for Pauli blocking and $+$ for Bose enhancement, and $d_k$ describes the degeneracy of the thermal parton.

One way to calculate the transition rate from Eq.\,(\ref{tr}) is to follow the energy loss calculation of Braaten and Thoma \cite{Braaten:1991jj,Braaten:1991we} by shifting the $\mathbf{k}^\prime$ integration to the exchanged momentum $\mathbf{q}=\mathbf{p}-\mathbf{p}^\prime=\mathbf{k}^\prime-\mathbf{k}$, and separately computing the contributions from soft $\sim gT$ and hard $\sim \sqrt{ET}$ momentum exchange. For that matter, in \cite{Braaten:1991jj,Braaten:1991we} an intermediate separation momentum $q^*$ was introduced, which dropped out in the final result, when adding the hard and soft contribution of the energy loss rate.

Unfortunately, in the calculation of the transition rates, $q^*$ will not drop out in the final result, because some necessary approximations and simplifications due to symmetries of the integrand that could be done in the calculation of the energy loss rate cannot be done in this case. 
For example, to maintain detailed balance, the exact expression $f(k)(1 - f(k^\prime))$ must be kept and, in particular, cannot be replaced by $(f(k)-f(k^\prime))/2$ as done in \cite{Braaten:1991jj}. In fact, this procedure would lead to negative transition rates.
Keeping the full expression $f(k)(1 - f(k^\prime))$ in the soft part of the calculation will not lead to a logarithmic divergence, and the above mentioned cancellation between the infrared divergence of the hard part and the ultraviolet divergence of the soft part will not occur. We will show below how this problem can be circumvented using the method of calculating collisional energy loss presented in \cite{Djordjevic:2006tw}.

For now, we separately compute the hard and the soft contribution introducing the separation scale $q^*$ and match both parts at an intermediate value of ${q^*}^2\sim \sqrt{ET} m_g$ (to be precise we use the $q^*$-value for which the analytic results for the soft and hard part of $dE/dt$ \cite{Braaten:1991jj,Braaten:1991we,Thomas:1991ea,GaleKapusta:1992} become equal).
$m_g^2=1/2(1+N_f/6)g^2T^2$ is the thermal gluon mass, the characteristic soft scale of the system. We vary $q^*$ around the mean value to get a feeling for the uncertainty that enters the result.
We refer to this method as `method A' in the following.

Eq.\,(\ref{tr}) can be rewritten by introducing the discussed shift of the integration to $q$ and also choosing $\mathbf{q}$ to lie on the $z$-axis, while placing $\mathbf{p}$ in the $x$-$z$-plane. Taking the limit $p\rightarrow \infty$ leads to 
\begin{align}\label{tr2}
 \frac{d\Gamma}{d\omega}~&(E,\omega,T)=\frac{d_k}{(2\pi)^3}\frac{1}{16\,E^2}\int_0^p dq \int_{\frac{q-\omega}{2}}^\infty dk\,\theta(q-|\omega|)\notag\\
&\times\int_0^{2\pi}\frac{d\phi_{kq|pq}}{2\pi}|\mathcal{M}|^2 f(k,T)(1\pm f(k^\prime,T))\,,
\end{align}
where $\phi_{pq|kq}$ is the angle between the $\mathbf{p}\times\mathbf{q}$ and the $\mathbf{k}\times\mathbf{q}$ plane. The integration limits and the $\theta$-function take care of the kinematic restrictions $q<k+k^\prime$, $q<p+p^\prime$ and $-q<\omega<q$.

For $E\gg T$ the scattering amplitude is dominated by $t$-channel exchange processes, for which the squared matrix elements read 
\begin{align}
    |\mathcal{M}|^2_{qq}&=\frac{4}{9}g^4\frac{s^2+u^2}{t^2}\,,~~~
    |\mathcal{M}|^2_{qg}=2 g^4\left(1-\frac{su}{t^2}\right)\,, \notag\\
    |\mathcal{M}|^2_{gq}&=2 g^4\left(1-\frac{su}{t^2}\right)\,,
    |\mathcal{M}|^2_{gg}=\frac{9}{2} g^4\left(\frac{17}{8}-\frac{su}{t^2}\right)\,,\label{matrixelements}
\end{align}
with the Mandelstam variables 
\begin{align}
    s&=-\frac{t}{2q^2}\left\{\left[(p+p^\prime)(k+k^\prime)+q^2\right]\right.\notag\\
    &~~~~~~~~~~~~~\left.-\cos(\phi_{pq|kq})\sqrt{(4 p p^\prime+t)(4 k k^\prime+t)}\right\}\,,\notag\\
    t&=\omega^2-q^2\,,\notag\\
    u&=-s-t\,.
\end{align}
$|\mathcal{M}|^2_{ij}$, where $i\in\{q,g\}$ is the squared matrix element for the process of a hard parton of type $i$ scattering with a thermal parton of type $j$.
Plugging these matrix elements into Eq.\,(\ref{tr2}) and solving the $k$- and $q$-integrations numerically after imposing $q^*$ as a lower cutoff on the $q$-integration, we obtain the hard part of the transition rate.
\begin{figure}[tb]
  \begin{center}
    \includegraphics[width=5cm]{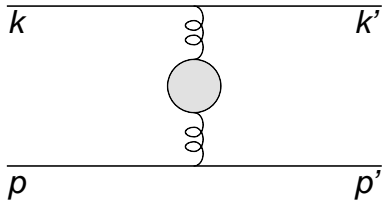}
    \caption{Feynman diagram for the amplitude that contributes to collisional energy loss in a QCD medium. The  hard thermal loop resummed gluon propagator is indicated by a gray blob.}
    \label{fig:feynman1}
  \end{center}
\end{figure}

To obtain the soft part, we can calculate the matrix element for e.g. $qq\rightarrow qq$ scattering 
from the diagram shown in Fig.\,\ref{fig:feynman1} using the effective thermal gluon propagator.
This method was used in \cite{Braaten:1991jj}, and further validation was put forward in \cite{Djordjevic:2006tw}.
The matrix element reads
\begin{align}
|\mathcal{M}|^2_{qq} = & \frac{8}{9} g^4 \textrm{Tr}({\not}P\gamma^\mu {\not}P^\prime\gamma^\nu) \textrm{Tr}({\not}K\gamma^\alpha
{\not}K^\prime\gamma^\beta)\notag\\ & \times D_{\mu\alpha}(Q) D^*_{\nu\beta}(Q)\,,
\end{align}
where four-momenta are denoted by capital letters, e.g. $Q=(\omega,\mathbf{q})$.
The effective thermal gluon propagator in the Coulomb gauge is given by
\begin{align}
    D^{\mu\nu}(Q)=\delta^{\mu 0}\delta^{\nu 0}\Delta_L(\omega,q)+P_T^{\mu\nu}\Delta_T(\omega,q)\,,
\end{align}
where $P_T^{00}=0,\,P_T^{ij}=\delta^{ij}-\hat{q}^i\hat{q}^j$ is the transverse projector, and
\begin{align}
    \Delta_L(\omega,q)&=\frac{-1}{q^2-m_g^2\left[x\ln\left(\frac{x+1}{x-1}\right)-2\right]}\,,\notag\\
    \Delta_T(\omega,q)&=\frac{-1}{q^2(x^2-1)-m_g^2\left[x^2+\frac{x}{2}(1-x^2)\ln\left(\frac{x+1}{x-1}\right)\right]}\notag\,,
\end{align}
are the longitudinal and transverse gluon propagators, with $x=\omega/q$. We find
\begin{align}\label{trexact}
 \int&\frac{d\phi_{kq|pq}}{2\pi}|\mathcal{M}|_{qq}^2=\notag\\
   &~~\frac{8}{9}g^4 p^2\Big\{[(k+k^\prime)^2-q^2]|\Delta_L|^2\notag\\
    &~~~~~~~~~~~+\frac{1}{2}\left(1-\frac{\omega^2}{q^2}\right)^2[(k+k^\prime)^2+q^2]|\Delta_T|^2\Big\}\,, 
\end{align}
which in the limit of small $\omega$ and $q$ becomes 
\begin{align}\label{trapp}
 \int\frac{d\phi_{kq|pq}}{2\pi}&|\mathcal{M}|_{qq}^2=\frac{32}{9}g^4 p^2 k k^\prime\notag\\
    &\times\left[|\Delta_L|^2+\frac{1}{2}\left(1-\frac{\omega^2}{q^2}\right)^2|\Delta_T|^2\right]\,.
\end{align}
Note that we keep a factor of $k^\prime=k+\omega$ even in this limit to fulfill detailed balance exactly.
Inserting this expression into Eq.\,(\ref{tr2}) and numerically solving the integrals after imposing $q^*$ as an upper cutoff on the $q$-integration, we obtain the final result for the soft part.
Results for the other processes follow analogously and only differ by the appropriate prefactors. Transition rates for gluon energy loss are obtained by multiplying those for quarks by $9/4$ \cite{Thomas:1991ea}.
The complete result is obtained by adding the hard and soft contributions.

Alternatively, we can use the same method as used for the soft part above and apply it to the whole transferred momentum range as done in \cite{Djordjevic:2006tw}. 
In this case we do not use approximations assuming small $\omega$ or $q$ as done in \cite{Braaten:1991jj} and our Eq.\,(\ref{trapp}), but insert Eq.\,(\ref{trexact}) directly into Eq.\,(\ref{tr2}).
Again, we solve the integrals in Eq.\,(\ref{tr2}) numerically and now obtain $q^*$-independent transition rates that
interpolate smoothly between the soft and hard regime of method A. 
We will refer to this method as `method B' in the following.
Note that because we are interested in the transition rate $d\Gamma/d\omega~(E,\omega,T)$ and not only in the integrated energy loss $dE/dt~(E,T)$, in Eq.\,(\ref{tr2}) we can not replace the Bose enhancement and Pauli blocking terms $(1\pm f(k^\prime,T))$ by $1$ as done in \cite{Djordjevic:2006tw}.

\begin{figure}[htb]
  \begin{center}
    \includegraphics[width=8cm]{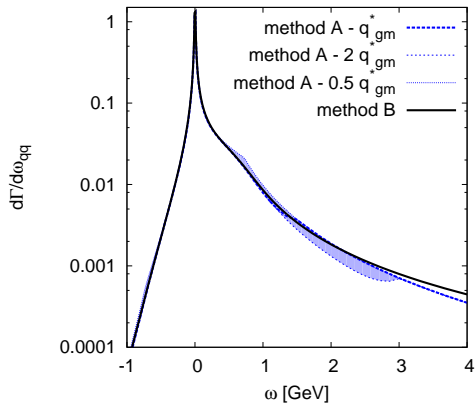}
    \caption{(Color online) Transition rate at $E=10\,\mathrm{GeV}$ and $T=200\,\mathrm{MeV}$ for the process $qq\rightarrow qq$ calculated in two different ways. $N_f=3$.}
    \label{fig:trqqMD}
  \end{center}
\end{figure}

In Fig.\,\ref{fig:trqqMD} we compare the transition rates computed using the two different methods described, and find a very good agreement between the two results as long as we choose $q^*$ to lie at the geometric mean between the hard $\sim \sqrt{ET}$ and the soft $\sim gT$ scale in method A. The band in Fig.\,\ref{fig:trqqMD} indicates the variation of the result when varying $q^*$ around that mean value by a factor of two in both directions. The slight difference for larger $\omega$ stems from the fact that in method B it was implicitly assumed that $-t\ll s$ and hence $s\approx -u$ (see \cite{Braaten:1991jj} and \cite{Thomas:1991ea} ), which was not done in our calculation of the hard part in method A (see Eq.\,(\ref{matrixelements})). We have verified that when using $s=-u$ in the calculation of the hard part in method A, the results from method A agree with those from method B at large $\omega$. The difference at large $\omega$ will be negligible for the jet evolution because elastic energy loss is dominated by soft momentum transfers.

As in \cite{Qin:2007rn}, we include the conversion processes (quark - anti-quark annihilation, pair production, and Compton scattering) that turn a quark into a gluon and vice versa.
There is no logarithmic enhancement of the u-channel exchange process as found for tagged heavy quark energy loss in \cite{Peigne:2007sd} and \cite{Peigne:2008nd}. See also \cite{Peigne:2008ns}. Also we neglect subleading constants arising from the u- and s-channel exchange. 
 
\section{Results}
We present results for the evolution of the momentum distribution of a parton with $E=10\,\mathrm{GeV}$, passing through ``a brick'' of QGP at constant temperatures $T=200\,\mathrm{MeV}$ and $T=300\,\mathrm{MeV}$, using $\alpha_s=g^2/(4\pi)=0.3$ and $N_f=3$. The values of these parameters were chosen as typical of the range in many phenomenological applications. Results for a parton with $E=100\,\mathrm{GeV}$ were found to be qualitatively very similar. We will show the distributions after the jet has passed through a 2 fm and 5 fm thick medium. 

\begin{figure}[htb]
  \begin{center}
    \includegraphics[width=9.5cm]{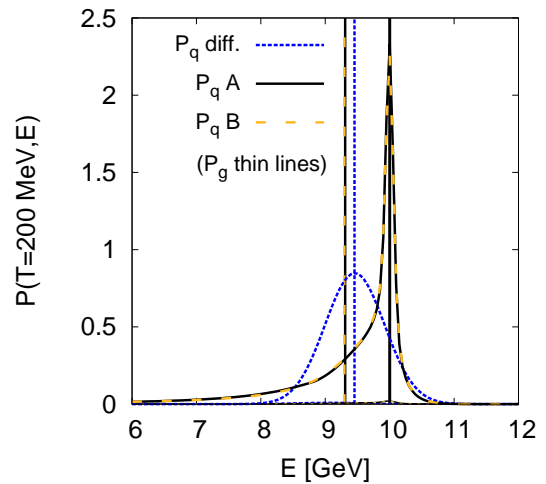}
    \caption{(Color online) Jet momentum distribution at $T=200\,\mathrm{MeV}$ for an initial quark jet with $E=10\,\mathrm{GeV}$ after 2 fm, for $\alpha_s=0.3$. The comparison shows the result for collisional energy loss obtained by the diffusion method and method A and B described above.}
    \label{fig:tqm20010-2fm-co}
  \end{center}
\end{figure}

\begin{figure}[htb]
  \begin{center}
    \includegraphics[width=9.5cm]{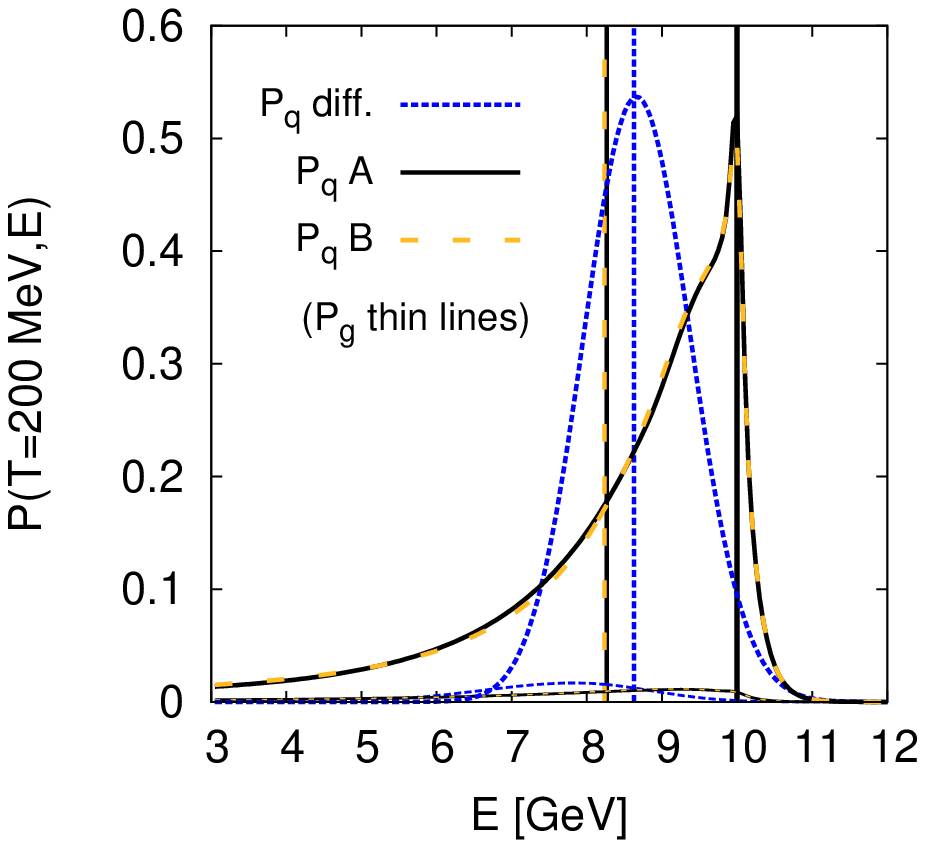}
    \caption{(Color online) Jet momentum distribution at $T=200\,\mathrm{MeV}$ for an initial quark jet with $E=10\,\mathrm{GeV}$ after 5 fm, for $\alpha_s=0.3$. 
    }
    \label{fig:tqm20010-5fm-co}
  \end{center}
\end{figure}

First we compare results from the diffusion method used in \cite{Qin:2007rn} to those obtained using methods A and B explained above for purely elastic energy loss.
Fig.\,\ref{fig:tqm20010-2fm-co} shows the momentum distributions after the quark has passed $2\,\mathrm{fm}$ of a $T=200\,\mathrm{MeV}$ plasma, Fig.\,\ref{fig:tqm20010-5fm-co} those after $5\,\mathrm{fm}$. The results obtained using methods A and B agree very well with each other but differ significantly from the Gaussian shape emerging using the diffusion method. The same qualitative result is obtained for a higher temperature $T=300\,\mathrm{MeV}$.

The total energy loss for both methods A and B is slightly larger than for the diffusion method. 
To understand this, let us consider the direct calculation of the energy loss rate.
In \cite{Romatschke:2003vc,Romatschke:2004au}, where the energy loss rate of a heavy fermion was calculated numerically and compared to the analytic result by Braaten and Thoma, it was found that the numerically calculated energy loss rate was usually larger than the analytic solution. 
Only for very small coupling the two results agreed around an intermediate $q^*$. 
The reason for the difference is that the analytic solution involves further approximations that allow both the hard and the soft contribution to become negative (which is necessary for the cancellation of $q^*$ in their sum), while the numerical result for both the hard and soft part is always positive. So the sum of both is generally larger than the sum of the analytic solutions.
Hence the difference in the mean energy loss is related to the uncertainty due to using perturbative methods at $\alpha_s=0.3$. Different approximations, all valid at infinitely small coupling, and hence all leading to the same result in this regime, can lead to different results when extrapolating to larger couplings.
Because in the diffusion method we use the analytic expressions for the energy loss rate but do not in the other methods, we find a qualitatively similar result to that of the direct energy loss rate calculation described above.

At $T=200\,\mathrm{MeV}$, the difference in the mean energy loss after $5\,\mathrm{fm}$ is approximately $30\%$ for purely collisional energy loss. It will be significantly smaller in the combined radiative and collisional calculation, because of the dominance of the radiative part. 
 
\begin{figure}[htb]
  \begin{center}
    \includegraphics[width=9.5cm]{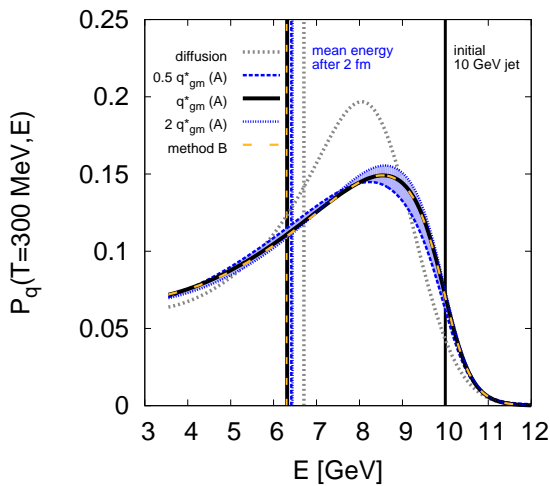}
    \caption{(Color online) $q^*$-dependence in method A and comparison to method B and the diffusion method. The result of method A using the geometric mean between the hard and the soft scale for $q^*$ coincides with the result obtained using method B.}
    \label{fig:qctqm}
  \end{center}
\end{figure}

Next, we show results including radiative energy loss. To get a feeling for how much the variation of $q^*$ in method A affects the final result, in Fig.\,\ref{fig:qctqm} we show the jet momentum distribution as found after the parton has passed a static $T=300\,\mathrm{MeV}$ QGP of length $2\,\mathrm{fm}$, using $q^*$ at the discussed mean value between the hard and soft momentum scale and for twice and half that value.
The band shown is a measure of the uncertainty related to the $q^*$-dependence in method A.

\begin{figure}[htb]
  \begin{center}
    \includegraphics[width=9.5cm]{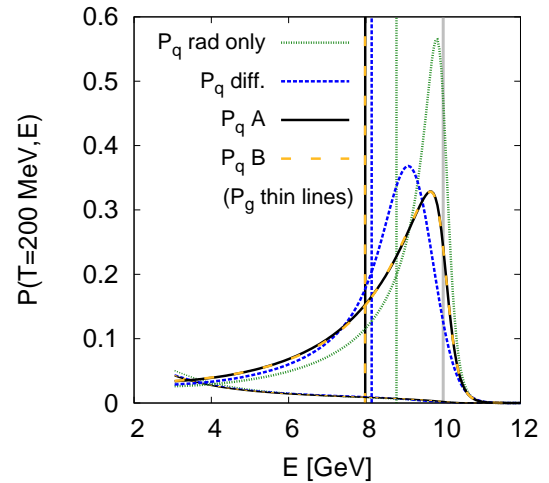}
    \caption{(Color online) Jet momentum distribution at $T=200\,\mathrm{MeV}$ for an initial quark jet with $E=10\,\mathrm{GeV}$ after 2 fm, for $\alpha_s=0.3$. The comparison shows the result for radiative energy loss only, collisional and radiative energy loss using the diffusion method, and the ones obtained by using method A and B described above.}
    \label{fig:tqm20010-2fm}
  \end{center}
\end{figure}

\begin{figure}[htb]
  \begin{center}
    \includegraphics[width=9.5cm]{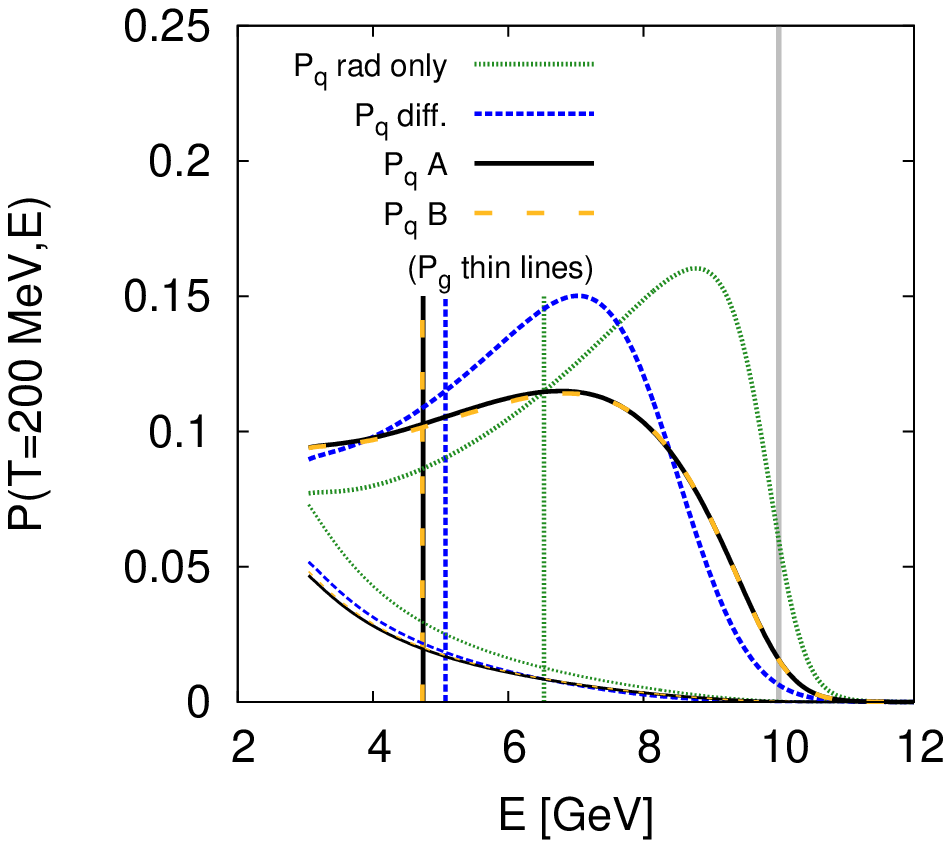}
    \caption{(Color online) Jet momentum distribution at $T=200\,\mathrm{MeV}$ for an initial quark jet with $E=10\,\mathrm{GeV}$ after 5 fm, for $\alpha_s=0.3$.
    }
    \label{fig:tqm20010-5fm}
  \end{center}
\end{figure}

\begin{figure}[htb]
  \begin{center}
    \includegraphics[width=9.5cm]{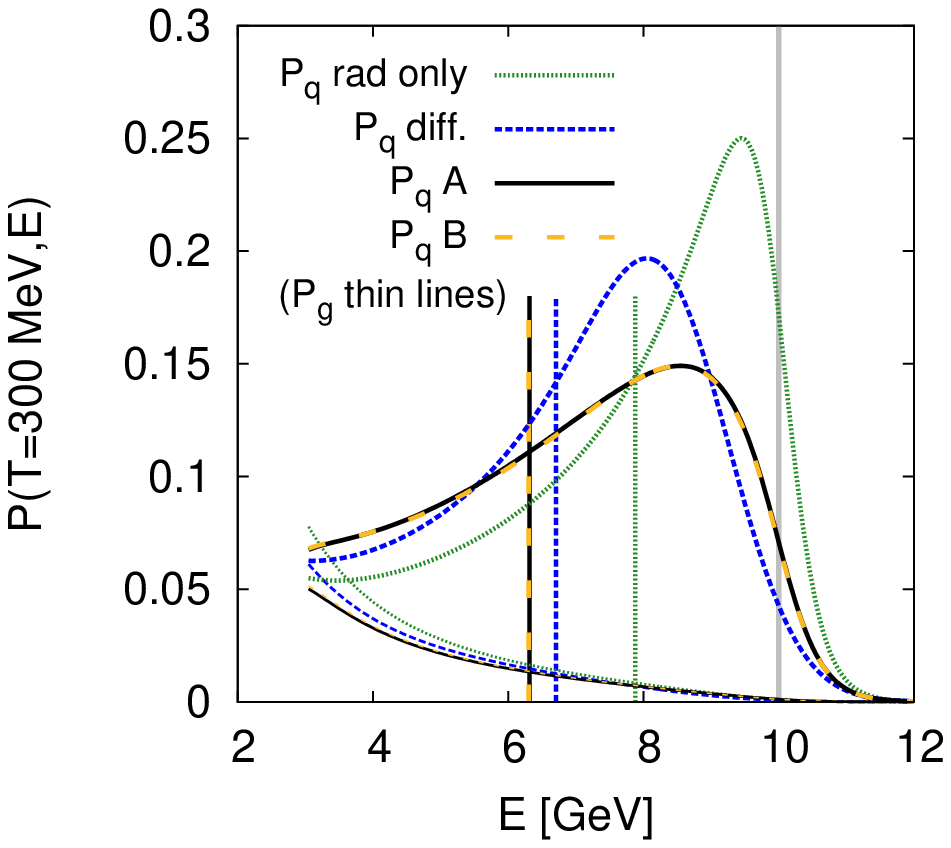}
    \caption{(Color online) Jet momentum distribution at $T=300\,\mathrm{MeV}$ for an initial quark jet with $E=10\,\mathrm{GeV}$ after 2 fm, for $\alpha_s=0.3$.
    }
    \label{fig:tqm30010-2fm}
  \end{center}
\end{figure}

\begin{figure}[htb]
  \begin{center}
    \includegraphics[width=9.5cm]{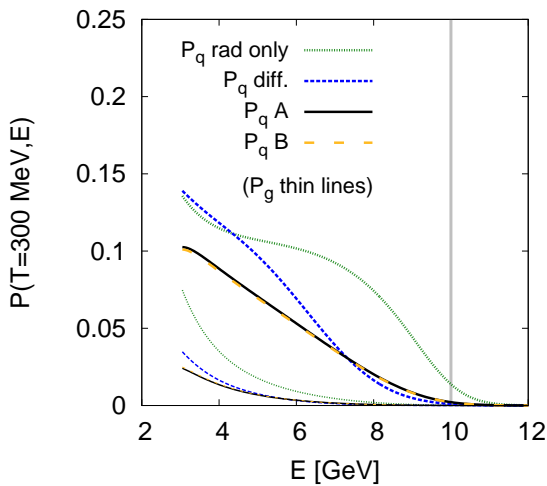}
    \caption{(Color online) Jet momentum distribution at $T=300\,\mathrm{MeV}$ for an initial quark jet with $E=10\,\mathrm{GeV}$ after 5 fm, for $\alpha_s=0.3$.
    }
    \label{fig:tqm30010-5fm}
  \end{center}
\end{figure}

Figs.\,\ref{fig:tqm20010-2fm} to \ref{fig:tqm30010-5fm} show the results including both radiative and collisional energy loss for two different temperatures and two different path lengths through the medium. The result for purely radiative energy loss is shown as well.
It can be seen that also in the combined radiative and collisional calculation the differences between the results obtained using the diffusion method and those obtained using method A or B are significant.

\section{Conclusion}
We studied the evolution of the momentum distribution of a high momentum parton traversing a brick of static QGP, considering radiative and collisional energy loss. For the collisional energy loss we used the leading order perturbative expression, which is an improvement over the method of approximating it by a mean energy loss accompanied by momentum diffusion, which has been used in recent works \cite{Wicks:2005gt,Qin:2007rn}.

While the mean energy loss is only slightly larger when using the more precise transition rates as opposed to the diffusion method, a significant difference in the final shape of the momentum distribution is found. It is thus entirely conceivable that, depending on the specific observables, this shape difference could play a large role in the interpretation of experimental measurements. Work in this direction is ongoing. 

We therefore conclude that it is important to go beyond the diffusion approximation, particularly
when comparing different theoretical energy loss formalisms including both radiative and collisional energy loss.  


\section*{Acknowledgments}
\hyphenation{Abhijit Majumder Ulrich Heinz Michael Strickland Richard Tomlinson}
We are happy to thank  Ulrich Heinz, Sangyong Jeon, Abhijit Majumder, and Michael Strickland for discussions. This work was supported in part by the Natural Sciences and Engineering Research Council of Canada, and in part by the U.S. Department of Energy
under Grant DE-FG02-01ER41190. B.S.\ gratefully acknowledges a Richard H.~Tomlinson Fellowship awarded by McGill University.
\bibliography{collamy}

\begin{thebibliography}{43}
\expandafter\ifx\csname natexlab\endcsname\relax\def\natexlab#1{#1}\fi
\expandafter\ifx\csname bibnamefont\endcsname\relax
  \def\bibnamefont#1{#1}\fi
\expandafter\ifx\csname bibfnamefont\endcsname\relax
  \def\bibfnamefont#1{#1}\fi
\expandafter\ifx\csname citenamefont\endcsname\relax
  \def\citenamefont#1{#1}\fi
\expandafter\ifx\csname url\endcsname\relax
  \def\url#1{\texttt{#1}}\fi
\expandafter\ifx\csname urlprefix\endcsname\relax\def\urlprefix{URL }\fi
\providecommand{\bibinfo}[2]{#2}
\providecommand{\eprint}[2][]{\url{#2}}

\bibitem[{\citenamefont{Adcox et~al.}(2002)}]{Adcox:2001jp}
\bibinfo{author}{\bibfnamefont{K.}~\bibnamefont{Adcox}} \bibnamefont{et~al.}
  (\bibinfo{collaboration}{PHENIX}), \bibinfo{journal}{Phys. Rev. Lett.}
  \textbf{\bibinfo{volume}{88}}, \bibinfo{pages}{022301}
  (\bibinfo{year}{2002}), \eprint{nucl-ex/0109003}.

\bibitem[{\citenamefont{Adler et~al.}(2002)}]{Adler:2002xw}
\bibinfo{author}{\bibfnamefont{C.}~\bibnamefont{Adler}} \bibnamefont{et~al.}
  (\bibinfo{collaboration}{STAR}), \bibinfo{journal}{Phys. Rev. Lett.}
  \textbf{\bibinfo{volume}{89}}, \bibinfo{pages}{202301}
  (\bibinfo{year}{2002}), \eprint{nucl-ex/0206011}.

\bibitem[{\citenamefont{Gyulassy and Wang}(1994)}]{Gyulassy:1993hr}
\bibinfo{author}{\bibfnamefont{M.}~\bibnamefont{Gyulassy}} \bibnamefont{and}
  \bibinfo{author}{\bibfnamefont{X.-n.} \bibnamefont{Wang}},
  \bibinfo{journal}{Nucl. Phys.} \textbf{\bibinfo{volume}{B420}},
  \bibinfo{pages}{583} (\bibinfo{year}{1994}), \eprint{nucl-th/9306003}.

\bibitem[{\citenamefont{Baier et~al.}(1997{\natexlab{a}})\citenamefont{Baier,
  Dokshitzer, Mueller, Peigne, and Schiff}}]{Baier:1996sk}
\bibinfo{author}{\bibfnamefont{R.}~\bibnamefont{Baier}},
  \bibinfo{author}{\bibfnamefont{Y.~L.} \bibnamefont{Dokshitzer}},
  \bibinfo{author}{\bibfnamefont{A.~H.} \bibnamefont{Mueller}},
  \bibinfo{author}{\bibfnamefont{S.}~\bibnamefont{Peigne}}, \bibnamefont{and}
  \bibinfo{author}{\bibfnamefont{D.}~\bibnamefont{Schiff}},
  \bibinfo{journal}{Nucl. Phys.} \textbf{\bibinfo{volume}{B484}},
  \bibinfo{pages}{265} (\bibinfo{year}{1997}{\natexlab{a}}),
  \eprint{hep-ph/9608322}.

\bibitem[{\citenamefont{Baier et~al.}(1998)\citenamefont{Baier, Dokshitzer,
  Mueller, and Schiff}}]{Baier:1998yf}
\bibinfo{author}{\bibfnamefont{R.}~\bibnamefont{Baier}},
  \bibinfo{author}{\bibfnamefont{Y.~L.} \bibnamefont{Dokshitzer}},
  \bibinfo{author}{\bibfnamefont{A.~H.} \bibnamefont{Mueller}},
  \bibnamefont{and} \bibinfo{author}{\bibfnamefont{D.}~\bibnamefont{Schiff}},
  \bibinfo{journal}{Phys. Rev.} \textbf{\bibinfo{volume}{C58}},
  \bibinfo{pages}{1706} (\bibinfo{year}{1998}), \eprint{hep-ph/9803473}.

\bibitem[{\citenamefont{Zakharov}(2001)}]{Zakharov:2000iz}
\bibinfo{author}{\bibfnamefont{B.~G.} \bibnamefont{Zakharov}},
  \bibinfo{journal}{JETP Lett.} \textbf{\bibinfo{volume}{73}},
  \bibinfo{pages}{49} (\bibinfo{year}{2001}), \eprint{hep-ph/0012360}.

\bibitem[{\citenamefont{Wang and Guo}(2001)}]{Wang:2001if}
\bibinfo{author}{\bibfnamefont{X.-N.} \bibnamefont{Wang}} \bibnamefont{and}
  \bibinfo{author}{\bibfnamefont{X.-f.} \bibnamefont{Guo}},
  \bibinfo{journal}{Nucl. Phys.} \textbf{\bibinfo{volume}{A696}},
  \bibinfo{pages}{788} (\bibinfo{year}{2001}), \eprint{hep-ph/0102230}.

\bibitem[{\citenamefont{Vitev and Gyulassy}(2002)}]{Vitev:2002pf}
\bibinfo{author}{\bibfnamefont{I.}~\bibnamefont{Vitev}} \bibnamefont{and}
  \bibinfo{author}{\bibfnamefont{M.}~\bibnamefont{Gyulassy}},
  \bibinfo{journal}{Phys. Rev. Lett.} \textbf{\bibinfo{volume}{89}},
  \bibinfo{pages}{252301} (\bibinfo{year}{2002}), \eprint{hep-ph/0209161}.

\bibitem[{\citenamefont{Jeon and Moore}(2005)}]{Jeon:2003gi}
\bibinfo{author}{\bibfnamefont{S.}~\bibnamefont{Jeon}} \bibnamefont{and}
  \bibinfo{author}{\bibfnamefont{G.~D.} \bibnamefont{Moore}},
  \bibinfo{journal}{Phys. Rev.} \textbf{\bibinfo{volume}{C71}},
  \bibinfo{pages}{034901} (\bibinfo{year}{2005}), \eprint{hep-ph/0309332}.

\bibitem[{\citenamefont{Salgado and Wiedemann}(2003)}]{Salgado:2003gb}
\bibinfo{author}{\bibfnamefont{C.~A.} \bibnamefont{Salgado}} \bibnamefont{and}
  \bibinfo{author}{\bibfnamefont{U.~A.} \bibnamefont{Wiedemann}},
  \bibinfo{journal}{Phys. Rev.} \textbf{\bibinfo{volume}{D68}},
  \bibinfo{pages}{014008} (\bibinfo{year}{2003}), \eprint{hep-ph/0302184}.

\bibitem[{\citenamefont{Majumder et~al.}(2007)\citenamefont{Majumder, Wang, and
  Wang}}]{Majumder:2004pt}
\bibinfo{author}{\bibfnamefont{A.}~\bibnamefont{Majumder}},
  \bibinfo{author}{\bibfnamefont{E.}~\bibnamefont{Wang}}, \bibnamefont{and}
  \bibinfo{author}{\bibfnamefont{X.-N.} \bibnamefont{Wang}},
  \bibinfo{journal}{Phys. Rev. Lett.} \textbf{\bibinfo{volume}{99}},
  \bibinfo{pages}{152301} (\bibinfo{year}{2007}), \eprint{nucl-th/0412061}.

\bibitem[{\citenamefont{Wicks et~al.}(2007)\citenamefont{Wicks, Horowitz,
  Djordjevic, and Gyulassy}}]{Wicks:2005gt}
\bibinfo{author}{\bibfnamefont{S.}~\bibnamefont{Wicks}},
  \bibinfo{author}{\bibfnamefont{W.}~\bibnamefont{Horowitz}},
  \bibinfo{author}{\bibfnamefont{M.}~\bibnamefont{Djordjevic}},
  \bibnamefont{and} \bibinfo{author}{\bibfnamefont{M.}~\bibnamefont{Gyulassy}},
  \bibinfo{journal}{Nucl. Phys.} \textbf{\bibinfo{volume}{A784}},
  \bibinfo{pages}{426} (\bibinfo{year}{2007}), \eprint{nucl-th/0512076}.

\bibitem[{\citenamefont{Zhang et~al.}(2007)\citenamefont{Zhang, Owens, Wang,
  and Wang}}]{Zhang:2007ja}
\bibinfo{author}{\bibfnamefont{H.}~\bibnamefont{Zhang}},
  \bibinfo{author}{\bibfnamefont{J.~F.} \bibnamefont{Owens}},
  \bibinfo{author}{\bibfnamefont{E.}~\bibnamefont{Wang}}, \bibnamefont{and}
  \bibinfo{author}{\bibfnamefont{X.-N.} \bibnamefont{Wang}},
  \bibinfo{journal}{Phys. Rev. Lett.} \textbf{\bibinfo{volume}{98}},
  \bibinfo{pages}{212301} (\bibinfo{year}{2007}), \eprint{nucl-th/0701045}.

\bibitem[{\citenamefont{Qin et~al.}(2007)}]{Qin:2007zz}
\bibinfo{author}{\bibfnamefont{G.-Y.} \bibnamefont{Qin}} \bibnamefont{et~al.},
  \bibinfo{journal}{Phys. Rev.} \textbf{\bibinfo{volume}{C76}},
  \bibinfo{pages}{064907} (\bibinfo{year}{2007}), \eprint{arXiv:0705.2575}.

\bibitem[{\citenamefont{Qin et~al.}(2008)}]{Qin:2007rn}
\bibinfo{author}{\bibfnamefont{G.-Y.} \bibnamefont{Qin}} \bibnamefont{et~al.},
  \bibinfo{journal}{Phys. Rev. Lett.} \textbf{\bibinfo{volume}{100}},
  \bibinfo{pages}{072301} (\bibinfo{year}{2008}), \eprint{arXiv:0710.0605}.

\bibitem[{\citenamefont{Fochler et~al.}(2008)\citenamefont{Fochler, Xu, and
  Greiner}}]{Fochler:2008ts}
\bibinfo{author}{\bibfnamefont{O.}~\bibnamefont{Fochler}},
  \bibinfo{author}{\bibfnamefont{Z.}~\bibnamefont{Xu}}, \bibnamefont{and}
  \bibinfo{author}{\bibfnamefont{C.}~\bibnamefont{Greiner}}
  (\bibinfo{year}{2008}), \eprint{arXiv:0806.1169}.

\bibitem[{\citenamefont{Schenke et~al.}(2009)\citenamefont{Schenke, Strickland,
  Dumitru, Nara, and Greiner}}]{Schenke:2008gg}
\bibinfo{author}{\bibfnamefont{B.}~\bibnamefont{Schenke}},
  \bibinfo{author}{\bibfnamefont{M.}~\bibnamefont{Strickland}},
  \bibinfo{author}{\bibfnamefont{A.}~\bibnamefont{Dumitru}},
  \bibinfo{author}{\bibfnamefont{Y.}~\bibnamefont{Nara}}, \bibnamefont{and}
  \bibinfo{author}{\bibfnamefont{C.}~\bibnamefont{Greiner}},
  \bibinfo{journal}{Phys. Rev.} \textbf{\bibinfo{volume}{C79}},
  \bibinfo{pages}{034903} (\bibinfo{year}{2009}), \eprint{arXiv:0810.1314}.

\bibitem[{\citenamefont{Baier et~al.}(1997{\natexlab{b}})\citenamefont{Baier,
  Dokshitzer, Mueller, Peigne, and Schiff}}]{Baier:1996kr}
\bibinfo{author}{\bibfnamefont{R.}~\bibnamefont{Baier}},
  \bibinfo{author}{\bibfnamefont{Y.~L.} \bibnamefont{Dokshitzer}},
  \bibinfo{author}{\bibfnamefont{A.~H.} \bibnamefont{Mueller}},
  \bibinfo{author}{\bibfnamefont{S.}~\bibnamefont{Peigne}}, \bibnamefont{and}
  \bibinfo{author}{\bibfnamefont{D.}~\bibnamefont{Schiff}},
  \bibinfo{journal}{Nucl. Phys.} \textbf{\bibinfo{volume}{B483}},
  \bibinfo{pages}{291} (\bibinfo{year}{1997}{\natexlab{b}}),
  \eprint{hep-ph/9607355}.

\bibitem[{\citenamefont{Gyulassy et~al.}(2001)\citenamefont{Gyulassy, Levai,
  and Vitev}}]{Gyulassy:2000er}
\bibinfo{author}{\bibfnamefont{M.}~\bibnamefont{Gyulassy}},
  \bibinfo{author}{\bibfnamefont{P.}~\bibnamefont{Levai}}, \bibnamefont{and}
  \bibinfo{author}{\bibfnamefont{I.}~\bibnamefont{Vitev}},
  \bibinfo{journal}{Nucl. Phys.} \textbf{\bibinfo{volume}{B594}},
  \bibinfo{pages}{371} (\bibinfo{year}{2001}), \eprint{nucl-th/0006010}.

\bibitem[{\citenamefont{Kovner and Wiedemann}(2003)}]{Kovner:2003zj}
\bibinfo{author}{\bibfnamefont{A.}~\bibnamefont{Kovner}} \bibnamefont{and}
  \bibinfo{author}{\bibfnamefont{U.~A.} \bibnamefont{Wiedemann}},
  \bibinfo{journal}{Review for Quark Gluon Plasma 3, Editors: R.C. Hwa and X.N.
  Wang, World Scientific, Singapore, 192}  (\bibinfo{year}{2003}),
  \eprint{hep-ph/0304151}.

\bibitem[{\citenamefont{Zakharov}(2007)}]{Zakharov:2007pj}
\bibinfo{author}{\bibfnamefont{B.~G.} \bibnamefont{Zakharov}},
  \bibinfo{journal}{JETP Lett.} \textbf{\bibinfo{volume}{86}},
  \bibinfo{pages}{444} (\bibinfo{year}{2007}), \eprint{0708.0816}.

\bibitem[{\citenamefont{Arnold et~al.}(2001{\natexlab{a}})\citenamefont{Arnold,
  Moore, and Yaffe}}]{Arnold:2001ms}
\bibinfo{author}{\bibfnamefont{P.}~\bibnamefont{Arnold}},
  \bibinfo{author}{\bibfnamefont{G.~D.} \bibnamefont{Moore}}, \bibnamefont{and}
  \bibinfo{author}{\bibfnamefont{L.~G.} \bibnamefont{Yaffe}},
  \bibinfo{journal}{JHEP} \textbf{\bibinfo{volume}{12}}, \bibinfo{pages}{009}
  (\bibinfo{year}{2001}{\natexlab{a}}), \eprint{hep-ph/0111107}.

\bibitem[{\citenamefont{Bjorken}(1982)}]{Bjorken:1982tu}
\bibinfo{author}{\bibfnamefont{J.~D.} \bibnamefont{Bjorken}},
  \bibinfo{journal}{FERMILAB-PUB-82-059-THY}  (\bibinfo{year}{1982}).

\bibitem[{\citenamefont{Renk}(2007)}]{Renk:2007id}
\bibinfo{author}{\bibfnamefont{T.}~\bibnamefont{Renk}}, \bibinfo{journal}{Phys.
  Rev.} \textbf{\bibinfo{volume}{C76}}, \bibinfo{pages}{064905}
  (\bibinfo{year}{2007}), \eprint{arXiv:0708.4319}.

\bibitem[{\citenamefont{Mustafa and Thoma}(2005)}]{Mustafa:2003vh}
\bibinfo{author}{\bibfnamefont{M.~G.} \bibnamefont{Mustafa}} \bibnamefont{and}
  \bibinfo{author}{\bibfnamefont{M.~H.} \bibnamefont{Thoma}},
  \bibinfo{journal}{Acta Phys. Hung.} \textbf{\bibinfo{volume}{A22}},
  \bibinfo{pages}{93} (\bibinfo{year}{2005}), \eprint{hep-ph/0311168}.

\bibitem[{\citenamefont{Mustafa}(2005)}]{Mustafa:2004dr}
\bibinfo{author}{\bibfnamefont{M.~G.} \bibnamefont{Mustafa}},
  \bibinfo{journal}{Phys. Rev.} \textbf{\bibinfo{volume}{C72}},
  \bibinfo{pages}{014905} (\bibinfo{year}{2005}), \eprint{hep-ph/0412402}.

\bibitem[{\citenamefont{Adil et~al.}(2007)\citenamefont{Adil, Gyulassy,
  Horowitz, and Wicks}}]{Adil:2006ei}
\bibinfo{author}{\bibfnamefont{A.}~\bibnamefont{Adil}},
  \bibinfo{author}{\bibfnamefont{M.}~\bibnamefont{Gyulassy}},
  \bibinfo{author}{\bibfnamefont{W.~A.} \bibnamefont{Horowitz}},
  \bibnamefont{and} \bibinfo{author}{\bibfnamefont{S.}~\bibnamefont{Wicks}},
  \bibinfo{journal}{Phys. Rev.} \textbf{\bibinfo{volume}{C75}},
  \bibinfo{pages}{044906} (\bibinfo{year}{2007}), \eprint{nucl-th/0606010}.

\bibitem[{\citenamefont{Wicks and Gyulassy}(2007)}]{Wicks:2007mk}
\bibinfo{author}{\bibfnamefont{S.}~\bibnamefont{Wicks}} \bibnamefont{and}
  \bibinfo{author}{\bibfnamefont{M.}~\bibnamefont{Gyulassy}},
  \bibinfo{journal}{J. Phys.} \textbf{\bibinfo{volume}{G34}},
  \bibinfo{pages}{S989} (\bibinfo{year}{2007}), \eprint{nucl-th/0701088}.

\bibitem[{\citenamefont{Majumder}(2008)}]{Majumder:2008zg}
\bibinfo{author}{\bibfnamefont{A.}~\bibnamefont{Majumder}}
  (\bibinfo{year}{2008}), \eprint{arXiv:0810.4967}.

\bibitem[{\citenamefont{Wang}(2007)}]{Wang:2006qr}
\bibinfo{author}{\bibfnamefont{X.-N.} \bibnamefont{Wang}},
  \bibinfo{journal}{Phys. Lett.} \textbf{\bibinfo{volume}{B650}},
  \bibinfo{pages}{213} (\bibinfo{year}{2007}), \eprint{nucl-th/0604040}.

\bibitem[{\citenamefont{Arnold et~al.}(2001{\natexlab{b}})\citenamefont{Arnold,
  Moore, and Yaffe}}]{Arnold:2001ba}
\bibinfo{author}{\bibfnamefont{P.}~\bibnamefont{Arnold}},
  \bibinfo{author}{\bibfnamefont{G.~D.} \bibnamefont{Moore}}, \bibnamefont{and}
  \bibinfo{author}{\bibfnamefont{L.~G.} \bibnamefont{Yaffe}},
  \bibinfo{journal}{JHEP} \textbf{\bibinfo{volume}{11}}, \bibinfo{pages}{057}
  (\bibinfo{year}{2001}{\natexlab{b}}), \eprint{hep-ph/0109064}.

\bibitem[{\citenamefont{Arnold et~al.}(2002)\citenamefont{Arnold, Moore, and
  Yaffe}}]{Arnold:2002ja}
\bibinfo{author}{\bibfnamefont{P.}~\bibnamefont{Arnold}},
  \bibinfo{author}{\bibfnamefont{G.~D.} \bibnamefont{Moore}}, \bibnamefont{and}
  \bibinfo{author}{\bibfnamefont{L.~G.} \bibnamefont{Yaffe}},
  \bibinfo{journal}{JHEP} \textbf{\bibinfo{volume}{06}}, \bibinfo{pages}{030}
  (\bibinfo{year}{2002}), \eprint{hep-ph/0204343}.

\bibitem[{\citenamefont{Turbide et~al.}(2005)\citenamefont{Turbide, Gale, Jeon,
  and Moore}}]{Turbide:2005fk}
\bibinfo{author}{\bibfnamefont{S.}~\bibnamefont{Turbide}},
  \bibinfo{author}{\bibfnamefont{C.}~\bibnamefont{Gale}},
  \bibinfo{author}{\bibfnamefont{S.}~\bibnamefont{Jeon}}, \bibnamefont{and}
  \bibinfo{author}{\bibfnamefont{G.~D.} \bibnamefont{Moore}},
  \bibinfo{journal}{Phys. Rev.} \textbf{\bibinfo{volume}{C72}},
  \bibinfo{pages}{014906} (\bibinfo{year}{2005}), \eprint{hep-ph/0502248}.

\bibitem[{\citenamefont{Braaten and
  Thoma}(1991{\natexlab{a}})}]{Braaten:1991jj}
\bibinfo{author}{\bibfnamefont{E.}~\bibnamefont{Braaten}} \bibnamefont{and}
  \bibinfo{author}{\bibfnamefont{M.~H.} \bibnamefont{Thoma}},
  \bibinfo{journal}{Phys. Rev.} \textbf{\bibinfo{volume}{D44}},
  \bibinfo{pages}{1298} (\bibinfo{year}{1991}{\natexlab{a}}).

\bibitem[{\citenamefont{Braaten and
  Thoma}(1991{\natexlab{b}})}]{Braaten:1991we}
\bibinfo{author}{\bibfnamefont{E.}~\bibnamefont{Braaten}} \bibnamefont{and}
  \bibinfo{author}{\bibfnamefont{M.~H.} \bibnamefont{Thoma}},
  \bibinfo{journal}{Phys. Rev.} \textbf{\bibinfo{volume}{D44}},
  \bibinfo{pages}{2625} (\bibinfo{year}{1991}{\natexlab{b}}).

\bibitem[{\citenamefont{Thoma}(1991)}]{Thomas:1991ea}
\bibinfo{author}{\bibfnamefont{M.~H.} \bibnamefont{Thoma}},
  \bibinfo{journal}{Phys. Lett.} \textbf{\bibinfo{volume}{B273}},
  \bibinfo{pages}{128} (\bibinfo{year}{1991}).

\bibitem[{\citenamefont{Kapusta and Gale}(2006)}]{GaleKapusta:1992}
\bibinfo{author}{\bibfnamefont{J.~I.} \bibnamefont{Kapusta}} \bibnamefont{and}
  \bibinfo{author}{\bibfnamefont{C.}~\bibnamefont{Gale}},
  \emph{\bibinfo{title}{Finite-Temperature Field Theory: Principles and
  Applications}} (\bibinfo{publisher}{Cambridge University Press},
  \bibinfo{year}{2006}).

\bibitem[{\citenamefont{Djordjevic}(2006)}]{Djordjevic:2006tw}
\bibinfo{author}{\bibfnamefont{M.}~\bibnamefont{Djordjevic}},
  \bibinfo{journal}{Phys. Rev.} \textbf{\bibinfo{volume}{C74}},
  \bibinfo{pages}{064907} (\bibinfo{year}{2006}), \eprint{nucl-th/0603066}.

\bibitem[{\citenamefont{Peigne and
  Peshier}(2008{\natexlab{a}})}]{Peigne:2007sd}
\bibinfo{author}{\bibfnamefont{S.}~\bibnamefont{Peigne}} \bibnamefont{and}
  \bibinfo{author}{\bibfnamefont{A.}~\bibnamefont{Peshier}},
  \bibinfo{journal}{Phys. Rev.} \textbf{\bibinfo{volume}{D77}},
  \bibinfo{pages}{014015} (\bibinfo{year}{2008}{\natexlab{a}}),
  \eprint{0710.1266}.

\bibitem[{\citenamefont{Peigne and
  Peshier}(2008{\natexlab{b}})}]{Peigne:2008nd}
\bibinfo{author}{\bibfnamefont{S.}~\bibnamefont{Peigne}} \bibnamefont{and}
  \bibinfo{author}{\bibfnamefont{A.}~\bibnamefont{Peshier}},
  \bibinfo{journal}{Phys. Rev.} \textbf{\bibinfo{volume}{D77}},
  \bibinfo{pages}{114017} (\bibinfo{year}{2008}{\natexlab{b}}),
  \eprint{0802.4364}.

\bibitem[{\citenamefont{Peigne}(2008)}]{Peigne:2008ns}
\bibinfo{author}{\bibfnamefont{S.}~\bibnamefont{Peigne}}, \bibinfo{journal}{AIP
  Conf. Proc.} \textbf{\bibinfo{volume}{1038}}, \bibinfo{pages}{139}
  (\bibinfo{year}{2008}), \eprint{0806.0242}.

\bibitem[{\citenamefont{Romatschke and Strickland}(2004)}]{Romatschke:2003vc}
\bibinfo{author}{\bibfnamefont{P.}~\bibnamefont{Romatschke}} \bibnamefont{and}
  \bibinfo{author}{\bibfnamefont{M.}~\bibnamefont{Strickland}},
  \bibinfo{journal}{Phys. Rev.} \textbf{\bibinfo{volume}{D69}},
  \bibinfo{pages}{065005} (\bibinfo{year}{2004}), \eprint{hep-ph/0309093}.

\bibitem[{\citenamefont{Romatschke and Strickland}(2005)}]{Romatschke:2004au}
\bibinfo{author}{\bibfnamefont{P.}~\bibnamefont{Romatschke}} \bibnamefont{and}
  \bibinfo{author}{\bibfnamefont{M.}~\bibnamefont{Strickland}},
  \bibinfo{journal}{Phys. Rev.} \textbf{\bibinfo{volume}{D71}},
  \bibinfo{pages}{125008} (\bibinfo{year}{2005}), \eprint{hep-ph/0408275}.

\end{thebibliography}

\end{document}